\documentclass[twocolumn,showpacs,amsmath,amssymb,showkeys,aps,floatfix]{revtex4-1}
\usepackage{graphicx}
\usepackage{color}
\usepackage{float}
\usepackage[ps2pdf,bookmarks,urlcolor=blue,citecolor=blue,linkcolor=blue,
              pagecolor=blue,hyperindex,colorlinks,hyperfigures,
       ]{hyperref}
\begin{document}
\title{The role of correlations in the high-pressure phase of FeSe}
\author{
    S.~Shahab Naghavi,
    Stanislav Chadov,
    Claudia Felser}
\affiliation{$^1$Institut f\"ur Anorganische Chemie und
Analytische Chemie, Johannes Gutenberg - Universtit\"at,  55099
Mainz, Germany} 
\email{chadov@uni-mainz.de}
\date{\today}
\begin{abstract}
We present a systematic study of the high-pressure FeSe phase performed 
by means of the first-principle electronic structure calculations. Basing 
on available experimental information about the unit cell geometry we 
calculate the band structure and characterize the related properties during 
their pressure driven evolution. The electronic structure including the hybrid 
functional B3LYP or the Hubbard parameter $U$ for the iron $d$ states 
lead to the correct semiconducting ground state for the hexagonal stoichiometric 
FeSe within the broad pressure range (up to 30\,GPa).
\end{abstract}
\maketitle
The intriguing class of materials with respect to high-temperature 
superconductivity field is presented by iron arsenide based compounds~\cite{FeAS-layer1_FeSe,
FeAs-layer2_Fese,FeAS-layer3_FeSe,FeAs-layer-4_FeSe,FeAs-layer6_FeSe,
parker_structure_2009_FeSe} which exhibit the critical temperatures up 
to 55\,K~\cite{FeAs-layer5-55k_FeSe}. The study of these materials is complicated 
by the fact that their magnetic and superconducting states are competing 
at very similar conditions. For this reason the experiments under 
pressure~\cite{Our-nature_FeSe} gain special importance since the magnetism 
is typically suppressed by decreasing the volume. In addition, the pressure is 
a cleanest tuning parameter that allows to study the electronic structure by 
{\em ab-initio} methods.

\begin{figure}
\begin{center}
\includegraphics[width=1.0\linewidth,clip]{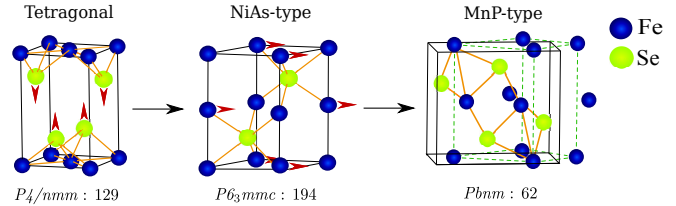}
\caption{Pressure driven evolution of FeSe phases: superconducting
  tetragonal ($Cmma$) phase is transformed to high-symmetry NiAs-type ($P6_3mmc$)
  structure, which in turn  transforms into lower-symmetry
  tetragonal MnP-type ($Pbnm$) phase by little distortion.} 
\label{FIG:fig1.eps}
\end{center}
\end {figure}

Probably the simplest material related to the iron arsenide family 
is the tetragonal ({\em P4/nmm}, group No.~129) $\beta$-phase of FeSe
formed by layers of edge-sharing tetrahedrons. Upon cooling the 
tetragonal phase undergoes a slight orthorhombic distortion by lowering 
its symmetry down to {\it Cmma} (Group No.~67)~\cite{FeSe-tetra-ortho-margadona_FeSe,FeSe-Ortho-Tetra_FeSe}. 
This phase transition takes place within a broad temperature range 
centered at about 100\,K depending on a crystal size and 
stoichiometry~\cite{FeSe-tetra-ortho-margadona_FeSe,FeSe-Ortho-Tetra_FeSe,Sensitivity-super_FeSe}. 
At rather ordinary temperature ($\sim 8$\,K) it becomes superconducting~\cite{FeSe_FeSe}, 
however, with pressure the $T_{\rm c}$ raises amazingly high (up to 
37\,K at about 7-9\,GPa)~\cite{FeSe1_FeSe,Our-nature_FeSe,NiAs-to-MnP-margadonna_FeSe}. 
This important information may indicate the direction to search for the new superconductors 
with even higher $T_{\rm c}$, which is the main focus of research.
At first glance, the pressure dependence of the $T_{\rm c}$ in FeSe is 
reminiscent of the superconducting dome observed in many unconventional 
superconductors, such as cuprates, heavy fermions and pnictides. 
However in contrast to these systems the vanishing of superconductivity 
in FeSe at very high pressure is related to a first-order structural phase 
transition to a hexagonal ({\it P6$_3$mmc}, NiAs-type) more densely packed
phase~\cite{Fe7Se8_Mag_FeSe,FeSe-Tc-34K_FeSe,Our-nature_FeSe} or its very similar 
low-temperature orthorhombic modification ({\em Pbnm}, MnP-type)~\cite{FeSe-Tc-34K_FeSe,Our-nature_FeSe}. 
The corresponding structures are shown in Figure~\ref{FIG:fig1.eps}. 

The small amount of NiAs-phase appears already at ambient 
pressures~\cite{Fe7Se8_Mag_FeSe,FeSe-NiAs-Paramagnetic_FeSe} and gradually substitutes the superconducting
{\it Cmma} phase until it fully converts into NiAs-type  at about 30\,GPa. 
This indicates that at least the  pressure range of superconducting dome 
could be extended if this structural phase transition is avoided. The central 
aspect in this direction is the knowledge of electronic structure of the 
NiAs-type phase and its related properties. Unfortunately there are no 
systematic studies on this system.

The crucial aspect for FeSe is the stoichiometry. Indeed, if the
tetragonal low-temperature Fe$_{1.01}$Se is superconducting, the small
increase of Fe amount (e.\,g. up to Fe$_{1.03}$Se)  already leads to a strong magnetic 
fluctuations which destroy the superconductivity~\cite{FeSe-Ortho-Tetra_FeSe,Stoicheometry-cava_FeSe}.
The analogous situation occurs by substituting Fe with small amount 
of Cu~\cite{Cu-dopp-FeSe_FeSe,Cu-Sub-FeSe_FeSe}. Many early experiments
 on non-stoichiometric hexagonal phases (e.\,g. Fe$_7$Se$_8$) performed
 at the ambient pressure report an antiferromagnetic 
order~\cite{Fe7Se8_Mag_FeSe,FeSe-NiAs-Paramagnetic_FeSe}. For thin films
the ferromagnetic order was reported as well~\cite{WZZ+08}.
On the other hand,  the stoichiometric hexagonal phase is unstable 
at ambient conditions however even for the lowest pressures the
magnetism reveals only in a form of dynamical fluctuations, but 
no net magnetic state was observed.

The first-principal description of the electronic structure under
pressure is a difficult task since it requires the adequate
``total energy--pressure'' mapping, or the so-called equation of state,
 based on a full structural optimization at each pressure point. This task is rather uncertain 
even for the high-symmetric structures with a single degree of freedom. 
Indeed, there are different uncontrolled sources of errors which become 
especially crucial for a certain pressure regime ranging from fundamental 
exchange-correlation inadequacies to the insufficient non-sphericity
of the one-particle potential.  

Present calculations are based on the so-called CRYSTAL06 code~\cite{Crystal06} 
which utilizes the local Gaussian basis well-suited for the description of the 
localized electrons in molecules. This formalism can be also applied
to describe the solid state systems, in particular
locally-correlated insulators and semiconductors. As a good choice for
the exchange-correlation potential within this formalism appears the so-called B3LYP 
functional~\cite{B3LYP} represented as the mixture of  LDA~\cite{VWN}, GGA~\cite{LYP} and the Hartree-Fock exact exchange.
Since the Hartree-Fock method systematically overestimates the band gap
and LDA  symmetrically underestimates it, the mixing coefficients are
found empirically in order to make use of error cancellation and to
improve the approach in average for the wide range of
systems. Since the B3LYP approach typically fails to describe itinerant magnetism,
 our calculations refer to the non-magnetic case. Fortunately, this
 corresponds to the experimental evidence which reports the absence of
magnetic order for the stoichiometric FeSe. 

In the following we perform the full optimization of geometry, i.\,e. without any
constraints for the lattice parameters and internal coordinates.  For
the starting values we use the experimental structural 
data~\cite{NiAs-to-MnP-margadonna_FeSe,Mnp-phase_FeSe}.
\begin{figure}[htp!]
\begin{center}
\includegraphics[width=1.0\linewidth]{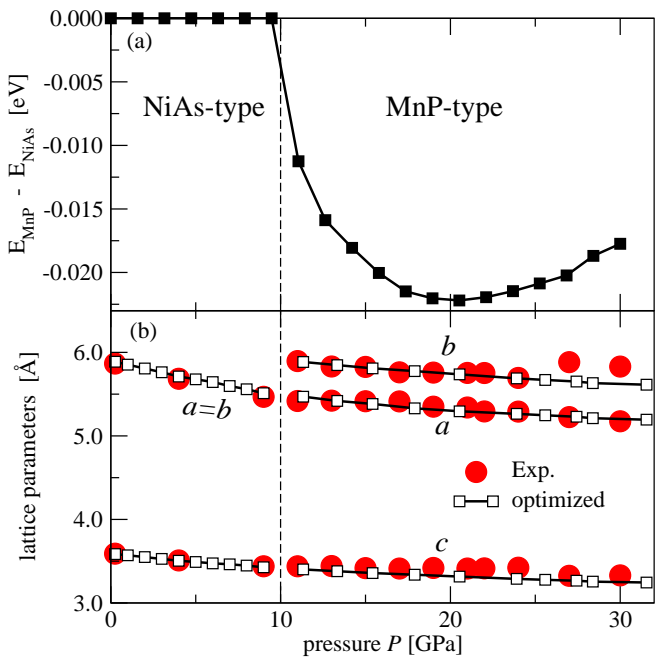} 
\caption{(color online) (a) Difference of the total energies (per atom) of
  NiAs- and MnP-type structures calculated as a function of pressure. (b) The pressure driven
  evolution of the corresponding lattice parameters. The results of
  optimization are displayed by squares. Red circles mark
  the experimental  data~\cite{NiAs-to-MnP-margadonna_FeSe,Mnp-phase_FeSe}. }
\label{FIG:fig2.eps}
\end{center}
\end {figure}
Figure~\ref{FIG:fig2.eps}\,(b)  demonstrates a very close agreement 
of the optimized structures for both MnP- and NiAs-type structures with experiment. 
As it follows from comparison of total energies (Figure~\ref{FIG:fig2.eps}\,(a)) the
high-symmetric NiAs-type structure is more stable  at low pressures,
while at high pressures the symmetry is reduced to more general MnP case. Since
this change is caused by a slight distortion (see Figure~\ref{FIG:fig2.eps}\,(b)) it does not
influence the electronic structure as indicated by comparison of the DOS
curves on Figure~\ref{FIG:fig3.eps}. For this reason in the
following we restrict  the consideration to more symmetric NiAs-type structure.

As it follows from Figure~\ref{FIG:fig3.eps},
\begin{figure}
\begin{center}
\includegraphics[width=1.0\linewidth]{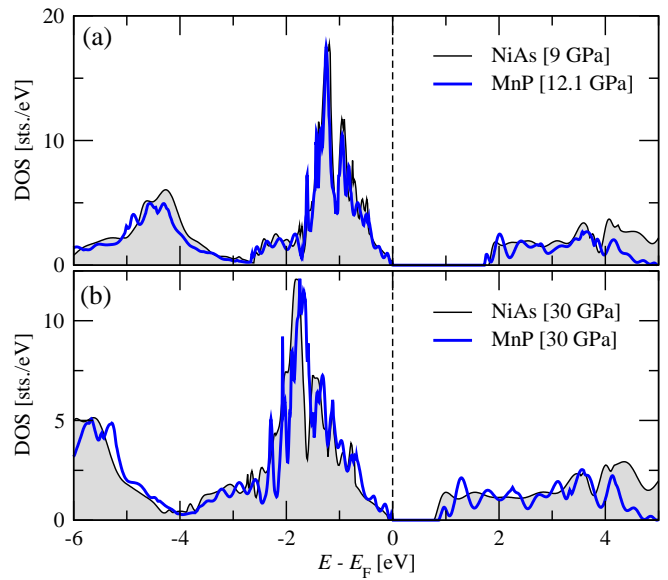}
\caption{(color online) The total DOS curves calculated for the
  optimized NiAs-type  (gray shaded area) and  MnP-type (blue solid line)  structures at low 
(a) and high (b) pressures.} 
\label{FIG:fig3.eps}
\end{center}
\end {figure}
NiAs-type (as well as MnP-type) phase exhibits a semiconducting
band gap.  This finding supports the experimental
reports~\cite{Our-nature_FeSe}. 
The calculated band gap width (about 2.5\,eV at ambient pressure) 
noticeably exceeds typical experimental values  (e.\,g. $\sim1.2$\,eV 
for the thin films of hexagonal
FeSe~\cite{FeSe-thin-film-gap-Hexa_FeSe}),  however such overestimation
is typical of all Hartree-Fock based methods.  

By increasing the pressure, the band gap gradually shrinks indicating
the possibility of insulator-metal transition at very high
pressures (above 30\,GPa). Such behavior suggests the
strongly-correlated origin of the band gap analogical to the
situation encountered e.\,g.  in transition metal oxides including the
known high-$T_{\rm c}$ superconductors~\cite{Pressure-Ind-Metallization_FeSe,Pressure-Ind-Metallization2_FeSe,LDA+U-Nonorthogonal_FeSe,
Oxychalcogenides+Antiferromagnetic_FeSe,What-Dom-Sc_FeSe} 
which exhibit the pressure driven insulator-metal transition accompanied 
by competition of the localized-itinerant electron contributions. 

\begin{figure}
\begin{center}
\includegraphics[width=1.0\linewidth]{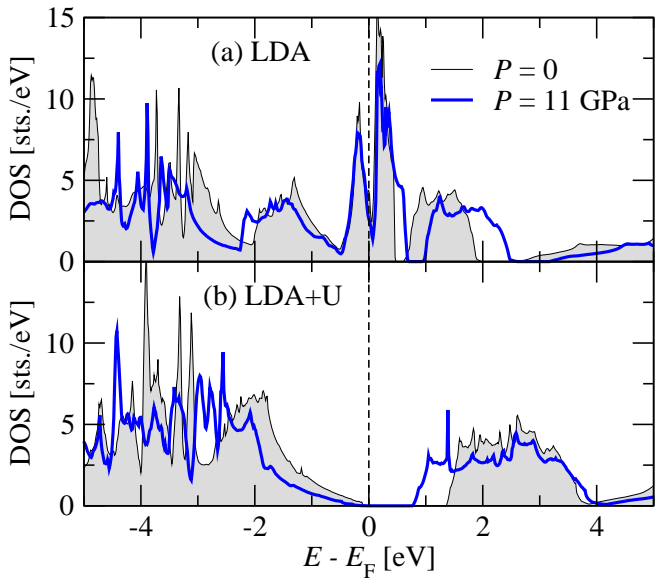}
\caption{(color online) 
	 DOS curves calculated at 0 (gray shaded area) and 11\,GPa (blue
         curve) for the NiAs-type phase using LMTO method. Panels (a) and (b) compare the plain
         LDA and LDA+U ($U_{\rm eff}=7$\,eV) results, respectively.} 
\label{FIG:fig4.eps}
\end{center}
\end {figure}

It is also instructive to probe the correlation-induced origin of the
gap  by applying the  alternative approaches which account  for the local correlations 
explicitly, as e.\,g. LDA+U method~\cite{LDA+U}.  On Figure~\ref{FIG:fig4.eps} 
we compare the LDA and LDA+U based total DOS curves calculated by 
LMTO (Local Muffin-Tin Orbitals) method~\cite{Anderson-LMTO} within the  so-called 
PY-LMTO package~\cite{PY-LMTO}.   Indeed, as it follows the band gap
could be explained on the basis of static approximation to the local on-site
electron correlation, as provided by LDA+U. 
At the same time the plain LDA description which lacks the proper treatment of strong correlation
delivers a metallic state in agreement with earlier
calculations~\cite{KZS+10}. The latter is characterized by a high peaks
of DOS at the Fermi energy, indicating the instability of the plain LDA solution. 
The band gap is opened due to the strong Coulomb repulsion which  splits the
correlated Fe $d$-shell into lower and upper Hubbard bands. 
By decreasing the volume the Hubbard bands broad, i.\,e. delocalize and the system
metallizes.

\begin{figure}
\begin{center}
\includegraphics[width=1.0\linewidth]{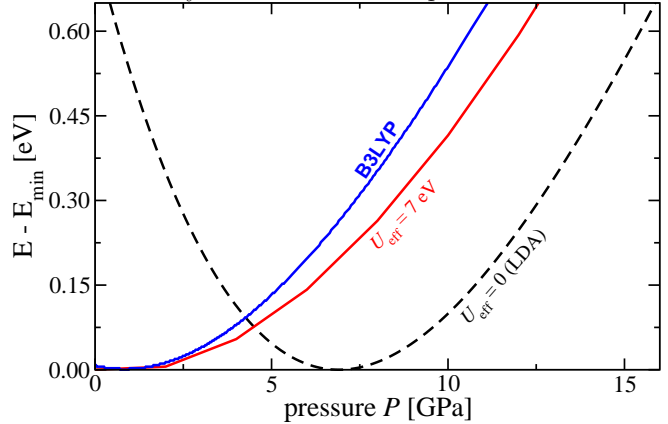}
\caption{(color online) 
	 Total energy of the NiAs-type phase calculated as a function of pressure
         within the LDA+U method for $U_{\rm eff}=0$ (plain LDA, dashed line) 
	 and $U_{\rm eff}=7$\,eV (red solid line), 
         compared to the B3LYP based  calculation
	 (blue solid line).} 
\label{FIG:fig5.eps}
\end{center}
\end {figure}

The central parameter of the theory is ${U_{\rm eff}=U-J}$, where $U$
and $J$ are the effective (screened) Coulomb direct 
and exchange interaction potentials. If the $J$ value can be relatively 
easy calculated from the first principles and typically does not exceed 1\,eV, 
the screened $U$ parameter makes problems. The bare  values of $U$ can be very 
high (about 10\,eV), however in  metals they are substantially scaled down 
 due to the intermediate  mobile electrons. The 
first-principle techniques to estimate the screening of $U$ parameter are 
 too imprecise and  computationally demanding in order to be used
as practical tool. In present case, since the geometry of the system is partially
known, we can estimate the adequate  $U_{\rm eff}$ from the following fit.
As it turns out from the pressure dependence of total energy  
(Figure~\ref{FIG:fig5.eps}), 
the plain LDA description ($U_{\rm eff}=0$) is indeed inadequate: 
the minimum of total energy is found at about 7\,GPa. 
By increasing $U_{\rm eff}$, the total energy minimum  shifts towards lower 
pressures and  finally riches the ambient  point at 
$U_{\rm eff}\approx7$\,eV. This huge value of $U$ is actually not much higher 
than typical values needed to obtain the adequate results in similar strongly
correlated systems~\cite{Mott-Insul-VO2_FeSe,Mutt-Insulator-hubbard_FeSe}.

Despite the noticeable difference of the band gap values (about 1.4\,eV
within LDA+U and 2.4\,eV within B3LYP), certain similarity between the 
Hartree-Fock and LDA+U exchange-correlation functionals lead to a
 similar behavior of the system under pressure which
indicates the Mott insulating origin of the NiAs-type phase. 

To conclude, we emphasize that the hexagonal phase of FeSe 
which substitutes  the tetragonal superconducting phase at higher
pressures  is characterized as locally correlated. These local 
correlations lead to an insulating state which can be classified as 
Mott insulator.  The band gap reduces due to increasing electron delocalization 
with pressure up to the insulator-metal transition which occurs 
roughly above 30\,GPa.  These results reasonably correspond to 
the experimental studies of FeSe and related systems under high pressure.

\begin{acknowledgments}
This work is part of the DFG priority program (SPP 1458)
“High-Temperature Superconductivity in Iron Pnictides”
funded by the German Science Foundation. The funding from the Graduate
School of Excellence Mainz is gratefully acknowledged.
\end{acknowledgments}

\end{document}